\documentstyle[aps,multicol]{revtex}
\def\narrowcaption{\columnwidth20.5pc}

\newlength{\defaultparindent}
\newenvironment{p-ni}{}{}

\newenvironment{Default Paragraph Font}{}{}

\begin{document}
\draft
\title{Detection of a Schr\"{o}dinger's Cat State in an rf-SQUID}
\author{Jonathan R. Friedman, Vijay Patel, W. Chen, S. K. Tolpygo \& J. E. Lukens}
\address{Department of Physics and Astronomy, The State University of New York at\\
Stony Brook, Stony Brook, NY 11794-3800}
\date{\today}
\maketitle

\begin{abstract}
We present experimental evidence for a coherent superposition of 
macroscopically distinct flux states in an rf-SQUID.  When the external 
flux $ \Phi_x $ applied to the SQUID is near 1/2 of a flux quantum $ \Phi_0 $,
the SQUID has two nearly degenerate configurations:  the zero- and one-fluxoid 
states, corresponding to a few microamperes of current flowing 
clockwise or counterclockwise, respectively.  The system is modeled as a 
particle in a double-well potential where each well represents a distinct 
fluxoid state (0 or 1) and the barrier between the wells can be controlled 
{\it in situ}. For low damping and a sufficiently high barrier, the system has 
a set of quantized energy levels localized in each well. The relative 
energies of these levels can be varied with $ \Phi_x $. External microwaves are
used to pump the system from the well-localized ground state of one well into 
one of a pair of excited states nearer the top of the barrier.  We 
spectroscopically map out the energy of these levels in the neighborhood of 
their degeneracy point by varying $\Phi_x$ as well as the barrier height.  
We find 
a splitting between the two states at this point, when both states 
are below the classical energy barrier, indicating that the system attains a 
coherent superposition of flux basis states that are macroscopically 
distinct in that their mean fluxes differ by more than 1/4 $\Phi_0 $ and their 
currents differ by several microamperes.
\end{abstract}

\smallskip

\begin{multicols}{2}
In 1935, Schr\"{o}dinger attempted to demonstrate the limitations of quantum
mechanics through a thought experiment in which a cat is put in a quantum
superposition of alive and dead states.\cite{SCat} The idea remained an
academic curiosity until the early 1980s when Leggett and co-workers\cite
{CaldeiraLeggett,LeggettRev,Weiss} proposed that under suitable conditions a
macroscopic object with many microscopic degrees of freedom could behave
quantum mechanically provided that it was sufficiently decoupled from its
environment. While much progress has been made in demonstrating the
macroscopic quantum behaviour of various systems such as superconductors, 
\cite{RousePRL,RouseRev,Clarke,Silvestrini,Nakamura} nanoscale magnets,\cite
{jrfMn12,delBarco,Wernsdorfer} laser-cooled trapped ions,\cite{Monroe}
photons in a microwave cavity\cite{Brune} and C$_{60}$ molecules,\cite{Arndt}
heretofore there has been no experimental demonstration of a quantum
superposition of truly macroscopically distinct states. Here we present the
first experimental evidence that a superconducting quantum interference
device (SQUID) can be put into a superposition of two magnetic-flux states,
one corresponding to a few microamperes of current flowing clockwise, the
other corresponding to the same amount of current flowing counterclockwise.

The basic rf-SQUID is a superconducting loop of inductance L broken by a
Josephson tunnel junction with capacitance C and critical current I$_{c}$.
Classically, when an external magnetic flux $\Phi _{x}$ is applied to the
SQUID, a superconducting (dissipationless) current will flow around the loop
to screen out the external flux. When the external flux reaches a certain
critical value (i.e. when the screening current reaches I$_{c}$), a flux
quantum $\Phi _{0}$ will enter the loop and the SQUID will again be in a
stable state. Quantum-mechanically, however, flux can tunnel into or out of
the SQUID before this critical value is reached. The dynamics (both
classical and quantum) of the SQUID are analogous to that of a particle of
mass C moving in a double-well potential given by
\begin{equation}
U=U_{0}\left[ {{\textstyle{\frac{1}{2}}}(\varphi -\varphi _{x})^{2}+\beta
_{L}\cos (\varphi )}\right],
\end{equation}
where $\varphi $ and $\varphi _{x}$ are the flux through the SQUID loop and
the external flux, respectively, in units of $\Phi _{0}/2\pi $ and measured
with respect to $\Phi _{0}/2$ ; $U_{0}\equiv \Phi _{0}^{2}/4\pi ^{2}L$ and $%
\beta _{L}\equiv 2\pi LI_{c}/\Phi _{0}$ characterize the energy barrier
between flux states. This potential is shown in Figure 1a, where the left
well corresponds to zero $\Phi _{0}$ in the SQUID and the right well one $%
\Phi _{0}$ . The system has energy levels localized in each well. When the
external flux is exactly $\Phi _{0}/2$ ($\varphi _{x}=0$), the potential is
symmetric. Any additional external flux tilts the potential, as shown in the
figure. At various values of external flux, levels in opposite wells will
align, giving rise to resonant tunnelling between flux states in the SQUID. 
\cite{RousePRL} However, until now, there has been no evidence that the
tunnelling process was coherent, that is, that the SQUID could be put into a
superposition of the two flux states.

Such a superposition would manifest itself in an anticrossing, as
illustrated in Figure 1b, where the energy level diagram of two levels of
different flux states (labelled $\left| 0\right\rangle $ and $\left|
1\right\rangle $) is shown in the neighbourhood in which they would become
degenerate without coherent interaction (dashed lines). Coherent tunnelling
lifts the degeneracy (solid lines) so that at the degeneracy point the
energy eigenstates are close to ${\textstyle{\frac{1}{\sqrt{2}}}}(\left|
0\right\rangle +\left| 1\right\rangle )$ and ${\textstyle{\frac{1}{\sqrt{2}}}%
}(\left| 0\right\rangle -\left| 1\right\rangle )$ , the symmetric and
antisymmetric superpositions.

The SQUID used in the experiments is made up of two Nb/AlOx/Nb tunnel
junctions in parallel, as shown in Fig. 1c; this essentially acts as a
tuneable junction in which I$_{c}$ can be adjusted with an applied flux $%
\Phi _{xdc}$ . Thus, with $\Phi _{x}$ we control the tilt $\varepsilon $ of
the potential in Figure 1a, while with $\Phi _{xdc}$ we control $\Delta
U_{0} $ , the height of the energy barrier at $\varepsilon $ =0. The flux
state of our sample is measured by a separate dc-SQUID magnetometer
inductively coupled to the sample. The sample is encased by a PdAu shield
that screens it from unwanted radiation; a coaxial cable entering the shield
allows the application of controlled external microwaves. The set up is
carefully filtered and shielded, as described elsewhere,\cite
{RousePRL,RouseRev} and cooled to 40 mK in a dilution refrigerator.

In our experiments, we probe the anticrossing of two excited levels in the
potential by using microwaves to produce photon-assisted tunnelling. Figure
1a depicts this process for the case where the levels $\left| 0\right\rangle 
$ and $\left| 1\right\rangle $ are each localized in opposite wells. The
system is initially placed in the lowest state in the left well (labelled $%
\left| i\right\rangle $) with the barrier high enough such that the
tunnelling rate is small. Microwave radiation is then applied. When the
energy difference between the initial state and an excited state matches the
radiation frequency, the system has an appreciable probability of making a
transition to the right well, which can be detected by the magnetometer.
Figure 2 shows schematically the photon-assisted process when the excited
levels have an anticrossing. Here the energy of the relevant levels is
plotted as a function of tilt $\varepsilon $. The dashed line represents
level $\left| i\right\rangle $ shifted upward by the energy of the
microwaves. At values of $\varepsilon $ for which this line intersects one
of the excited levels (indicated by the arrows), the system can absorb a
photon and make an interwell transition. When the barrier is reduced, the
excited levels move to lower energy (dotted lines in the figure) relative to 
$\left| i\right\rangle $ and photon absorption occurs at different values of 
$\varepsilon $. For a fixed frequency, we can map out the anticrossing by
progressively reducing the barrier and thus moving the levels through the
dashed photon line.

We use pulsed microwaves to excite the upper levels. Before each pulse, the
system is prepared in state $\left| i\right\rangle $ and the values of $%
\Delta U_{0}$ and $\varepsilon $ are set. Millisecond pulses of 96-GHz
microwave radiation at a fixed power are applied and the probability of
making a transition is measured. The experiment is repeated for various
values of $\varepsilon $ and $\Delta U_{0}$ . Data from these measurements
is shown in Figure 3, where the probability of making a photon-assisted
interwell transition is plotted as function of $\Phi _{x}$ . Each curve for
a given $\Delta U_{0}$ is shifted vertically for clarity. Two sets of peaks
are clearly seen. As $\Delta U_{0}$ is decreased, these peaks move closer
together and then move apart again. For $\Delta U_{0}$ = 9.117 K (red
curve), one peak is clearly higher than the other. Here the right peak
roughly corresponds to level $\left| 0\right\rangle $, which is localized is
the same well as $\left| i\right\rangle $. The matrix element for photon
absorption is larger for the $\left| i\right\rangle \rightarrow \left|
0\right\rangle $ transition than for the $\left| i\right\rangle \rightarrow
\left| 1\right\rangle $ transition, resulting in the asymmetry between the
peaks. When $\Delta U_{0}$ is decreased to 8.956 K (green curve), the peaks
move closer together and the asymmetry disappears. The two peaks now
approximately correspond to the coherent superpositions of the $\left|
0\right\rangle $ and $\left| i\right\rangle $ states; that is, ${\textstyle{%
\frac{1}{\sqrt{2}}}}(\left| 0\right\rangle +\left| 1\right\rangle )$ and ${%
\textstyle{\frac{1}{\sqrt{2}}}}(\left| 0\right\rangle -\left| 1\right\rangle
)$ . As the barrier is decreased further (8.797 K - violet curve), the peaks
move apart again and the asymmetry reappears, now with the left peak being
larger. The two levels have thus passed through the anticrossing, changing
roles without actually intersecting. The inset of the figure shows the
positions of the peaks in the main figure (as well as others peaks) in the $%
\Delta U_{0}$ - $\Phi _{x}$ plane. Two anticrossings are clearly seen. The
red lines are the results of a calculation and the violet line represents
the position of the classical energy barrier relative to state $\left|
i\right\rangle $. The position of this line relative to the data indicates
that both anticrossings correspond to levels that are below the barrier and
thus represent the superposition of macroscopically distinct states with
mean fluxes differing by about $\frac14$ $\Phi _{0}$ .

The peak positions in Figure 3 can also be used to calculate the level
energies. For one of the anticrossings, Figure 4 shows the energy of the
levels relative to the mean energy of $\left| 0\right\rangle $ and $\left|
1\right\rangle $ as a function of $\varepsilon $. The similarity of the data
to Figure 1b is now manifest. The figure also shows the results of a
calculation of the energy levels. At the middle of the anticrossing, the two
levels are separated by \~{}0.1 K in energy and the upper level is \~{}0.15 K
below the top of the classical energy barrier. There are three parameters
used for the calculations presented in Figures 3 and 4: L, L/C and $\beta
_{L}$, all of which can be independently determined from measurements of
classical phenomena or incoherent resonant tunnelling in the absence of
radiation. From independent measurements, we find L = 240$\pm $15 pH, L/C =
2300$\pm $10 $\Omega ^{2}$ and $\beta _{L}$ = 2.33$\pm $0.01. The values
used in the calculation that yielded the best agreement with the data are L
= 238 pH, L/C = 2300 $\Omega ^{2}$ and $\beta _{L}$ = 2.35, all in good
agreement with the independently determined values.

In closing, we would like to stress two related points regarding these
results. The first is that at the anticrossing both levels are below the top
of the classical energy barrier. This fact is essential for the system to be
in a superposition of macroscopically distinct states since the levels can
only be associated with one well (one flux state) if they are below the top
of the barrier. The second point concerns the meaning of ``macroscopic''.
The SQUID exhibits macroscopic quantum behaviour in two senses: 1) The
quantum dynamics of the SQUID is determined by the flux through the loop, a
collective coordinate representing the motion of \~{}10$^{9}$ Cooper pairs
acting in tandem. Since the experimental temperature is 1000 times smaller
than the superconducting energy gap, almost all microscopic degrees of
freedom are frozen out and only the collective flux coordinate retains any
dynamical relevance. 2) The two classical states that we find to be
superposed are macroscopically distinct. We calculate that for the
anticrossings measured, the states $\left| 0\right\rangle $ and $\left|
1\right\rangle $ differ in flux by more than $\frac14$ $\Phi _{0}$ and
differ in current by 2-3 microamperes. Given the SQUID's geometry, this
corresponds to a local magnetic moment of \~{}10$^{10}$ $\mu _{B} $, a truly
macroscopic moment.

We thank D. Averin and S. Han for many useful conversations, J. M\"{a}nnik
for technical assistance and M. P. Sarachik for the loan of some equipment.
This work was supported by the US Army Research Office and the US National
Science Foundation.

Correspondence and requests for materials should be addressed to J.R.F.
(e-mail: jonathan.friedman@sunysb.edu)

\begin{figure}[tbp]
\narrowcaption
\caption{{\bf a} SQUID potential. The left well corresponds to the zero-flux
state of the SQUID and the right well to the one-flux-quantum state. Energy
levels are localized in each well. Both the tilt $\varepsilon $ and energy
barrier $\Delta U_{0}$ can be varied {\it in situ} in the experiments. The process
of photon-induced interwell transitions is illustrated by the arrows, where
the system is excited out of the initial state $\left| i\right\rangle \ $and
into one of two excited states $\left| 0\right\rangle $ or $\left|
1\right\rangle $. {\bf b} Schematic anticrossing. When the two states $%
\left| 0\right\rangle $ and $\left| 1\right\rangle $ would classically
become degenerate, the degeneracy is lifted and the states of the system are
the symmetric and antisymmetric superpositions of the classical states: ${%
\textstyle{\frac{1}{\sqrt{2}}}}(\left| 0\right\rangle +\left| 1\right\rangle
)$ and ${\textstyle{\frac{1}{\sqrt{2}}}}(\left| 0\right\rangle -\left|
1\right\rangle )$ . {\bf c} Experimental set-up. Our SQUID contains a
``tuneable junction'', a small dc-SQUID. A flux $\Phi _{xdc}$ applied to
this small loop tunes the barrier height $\Delta U_{0}$ . Another flux $\Phi
_{x}$ tunes the tilt $\varepsilon $ of the potential. A separate SQUID acts
as a magnetometer, measuring the flux state of the sample.}
\end{figure}

\begin{figure}[tbp]
\narrowcaption
\caption{Illustration of experimental procedure. The applied microwaves
boost the system out of the initial state $\left| i\right\rangle $, bringing
it virtually to the dashed line. At certain values of $\varepsilon $ for
which this line intersects one of the excited states (indicated by the
arrows), a photon is absorbed and the system has a large probability of
making an interwell transition. When the energy barrier $\Delta U_{0}$ is
reduced, the levels move down relative to $\left| i\right\rangle $ (dotted
lines) and the values of $\varepsilon $ at which photon absorption occurs
changes.}
\end{figure}

\begin{figure}[tbp]
\narrowcaption
\caption{Experimental data. The main figure shows the probability of making
an interwell transition when a millisecond pulse of 96-GHz microwave
radiation is applied as a function of $\Phi _{x}$. For clarity, each curve
is shifted vertically by 0.3 relative to the previous one. As the energy
barrier is reduced, the two peaks observed move closer together and then
separate: the signature of an anticrossing. The inset shows the position of
the observed peaks in the $\Delta U_{0}$ - $\Phi _{x}$ plane. Also shown is
the calculated locus of points at which the virtual photon level (dashed
line in Figure 2) intersects an excited level (red lines) or the top of the
classical energy barrier (violet line).}
\end{figure}

\begin{figure}[tbp]
\narrowcaption
\caption{Energy of measured peaks relative to the calculated mean of the two
levels as a function of $\varepsilon $. At the midpoint of the figure, the
measured splitting between the two states in this anticrossing is \~{}0.1 K and
the upper level is \~{}0.15 K below the top of the classical energy barrier.
Calculated energy levels are also shown.}
\end{figure}

\end{multicols}

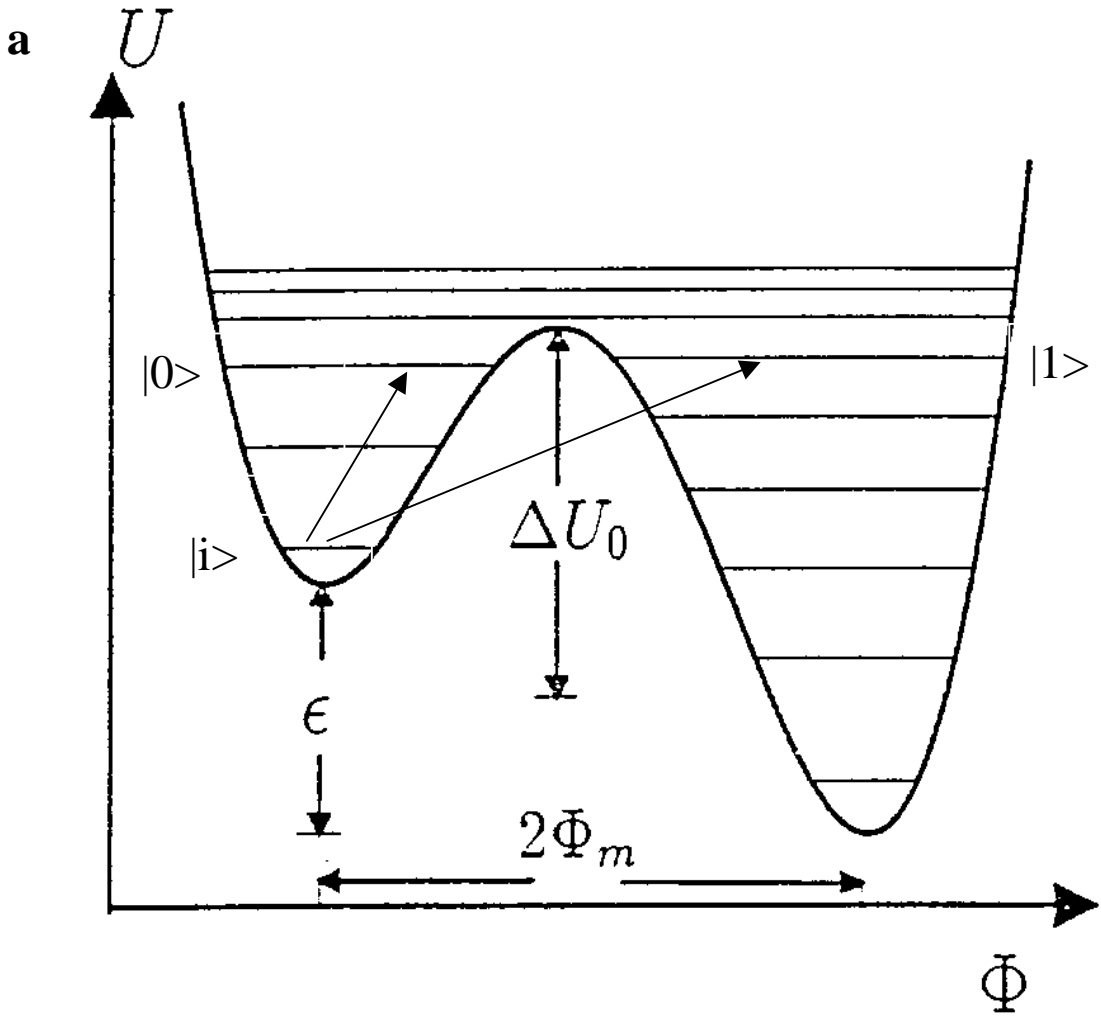
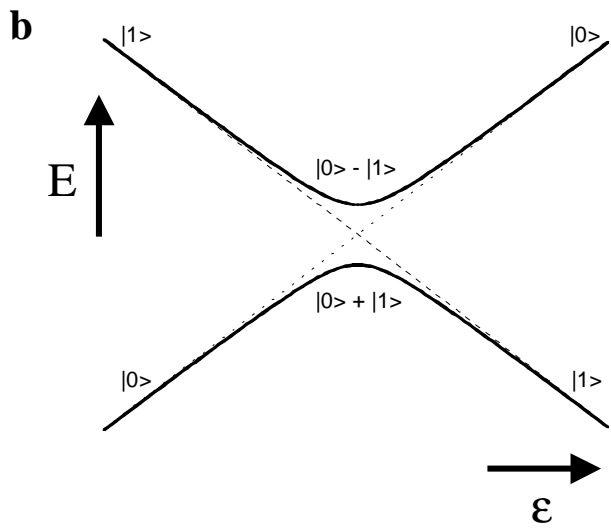
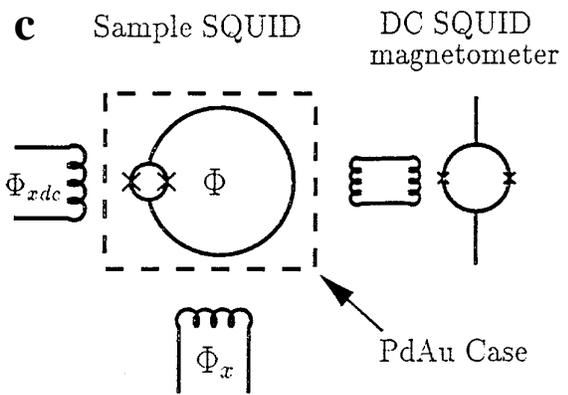
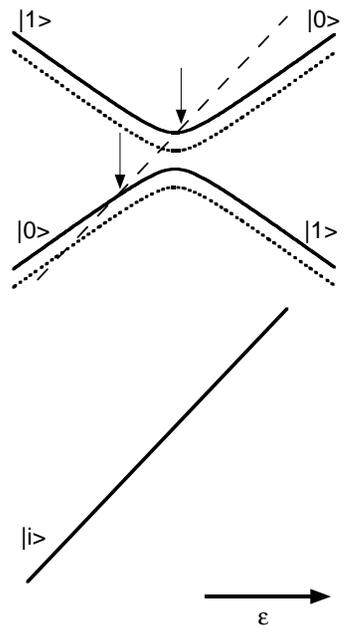
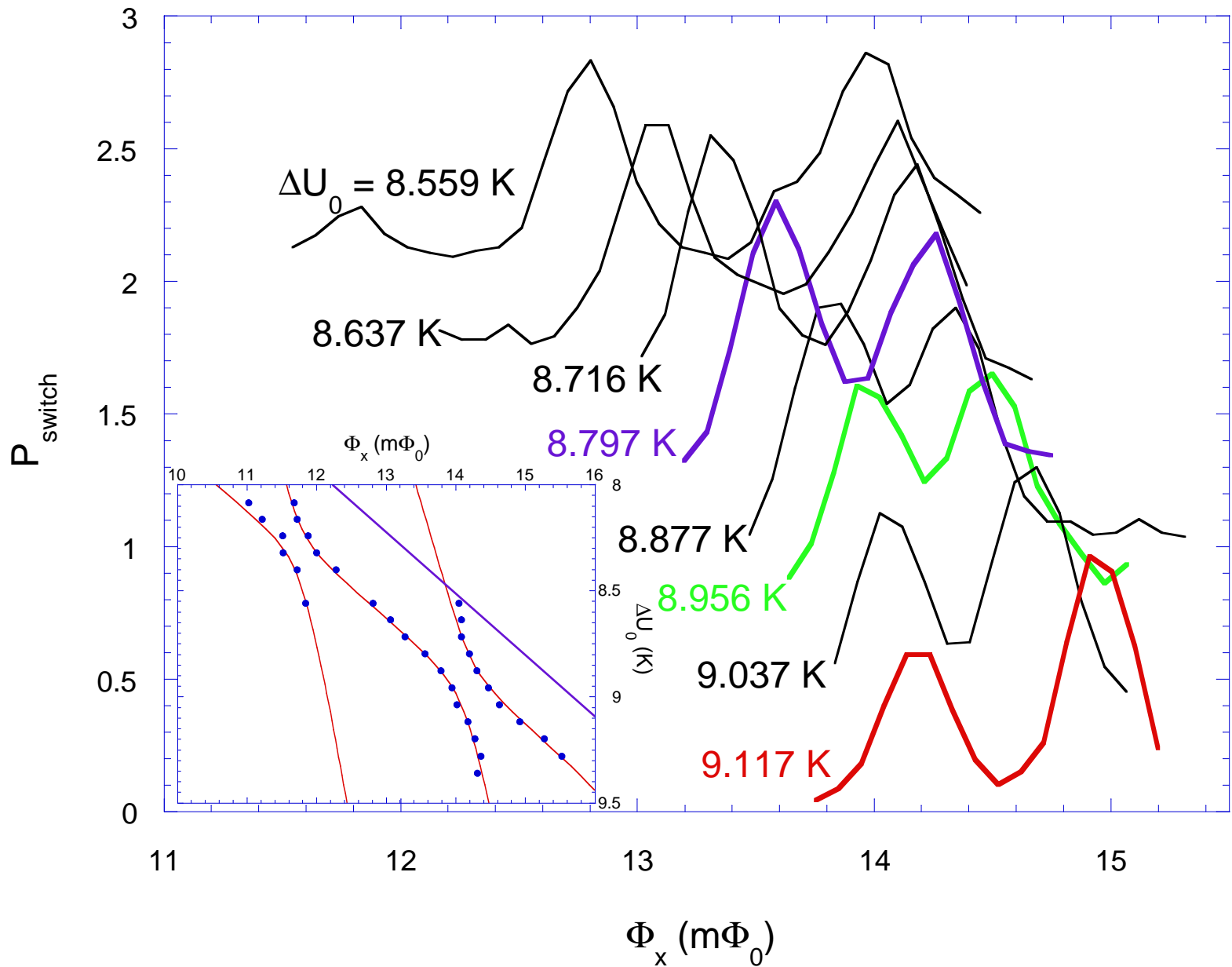
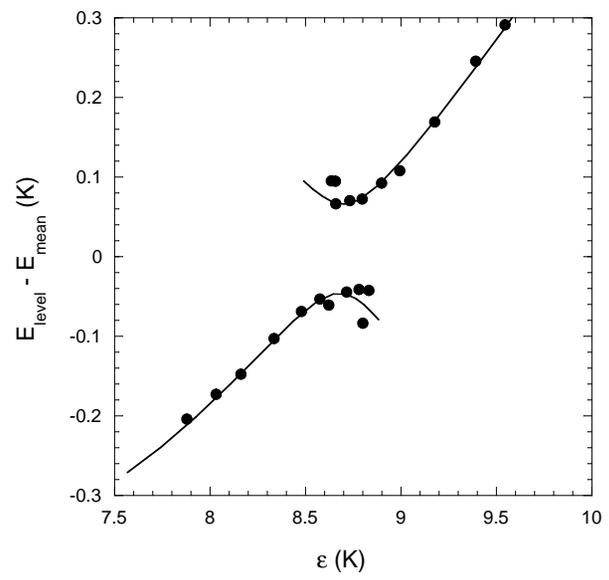
\end{document}